\documentclass[11pt]{article}
\usepackage[dvips]{graphicx}
\usepackage{amsmath,amssymb}
\usepackage{amscd}
\usepackage[all]{xy}

\newtheorem{Def}{Definition}
\newtheorem{Thm}[Def]{Theorem}
\newtheorem{Lem}[Def]{Lemma}
\newtheorem{Pro}[Def]{Proposition}

\begin{document}
\title{Permutation Complexity and Coupling Measures in Hidden Markov Models}

\author{Taichi Haruna\footnote{Corresponding author}$\ ^{\rm ,1}$,  Kohei Nakajima$\ ^{\rm 2}$ \\
\footnotesize{$\ ^{\rm 1}$ Department of Earth \& Planetary Sciences, Graduate School of Science, } \\
\footnotesize{Kobe University, 1-1, Rokkodaicho, Nada, Kobe 657-8501, Japan} \\
\footnotesize{$\ ^{\rm 2}$ Department of Mechanical and Process Engineering, } \\
\footnotesize{ETH Zurich, Leonhardstrasse 27, 8092 Zurich, Switzerland} \\
\footnotesize{E-mail: tharuna@penguin.kobe-u.ac.jp (T. Haruna)} \\
\footnotesize{Tel \& Fax: +81-78-803-5739} \\
}

\date{}
\maketitle

\begin{abstract}
In [Haruna, T. and Nakajima, K., 2011. Physica D 240, 1370-1377], the authors introduced 
the duality between values (words) and orderings (permutations) as a basis to discuss 
the relationship between information theoretic measures for finite-alphabet stationary 
stochastic processes and their permutation analogues. It has been used to give a simple 
proof of the equality between the entropy rate and the permutation entropy rate for 
any finite-alphabet stationary stochastic process and show some results on the excess 
entropy and the transfer entropy for finite-alphabet stationary ergodic Markov processes. 
In this paper, we extend our previous results to hidden Markov models and show the equalities between 
various information theoretic complexity and coupling measures and their permutation analogues. 
In particular, we show the following two results within the realm of hidden Markov models 
with ergodic internal processes: the two permutation analogues of the transfer entropy, 
the symbolic transfer entropy and the transfer entropy on rank vectors, 
are both equivalent to the transfer entropy if they are considered as the rates, 
and the directed information theory can be captured by the permutation entropy approach. 
\end{abstract}

{\bf Keywords:} Duality; Permutation Entropy; Excess Entropy; Transfer Entropy; Directed Information

\section{Introduction}
Recently, the permutation-information theoretic approach to time series analysis proposed by Bandt and Pompe \cite{Bandt2002a} 
has become popular in various fields \cite{Amigo2010}. It has been proved that the method of permutation is 
easy to implement relative to the other traditional methods and is robust under the existence of noise \cite{Bahraminasab2008,Cao2004,Kugiumtzis2013,Nakajima2013,Rosso2007}. 
However, if we turn our eyes to its theoretical side, few results are known for the permutation analogues of 
information theoretic measures except the entropy rate. 

There are two approaches to introduce permutation into dynamical systems theory. 
The first approach is introduced by Bandt et al. \cite{Bandt2002b}. Given a one-dimensional interval map, 
they considered permutations induced by iterations of the map. Each point in the interval 
is classified into one of $n!$ permutations according to the permutation defined by $n-1$ times 
iterations of the map starting from the point. Then, the Shannon entropy of this partition (called standard partition) 
of the interval is taken and normalized by $n$. The quantity obtained in the limit $n \to \infty$ 
is called permutation entropy if it exists. It was proved that the permutation 
entropy is equal to the Kolmogorov-Sinai entropy for any piecewise monotone interval map \cite{Bandt2002b}. 
This approach based on the standard partitions was extended by \cite{Keller2010, Keller2012}. 

The second approach is taken by Amig\'{o} et al. \cite{Amigo2005,Amigo2010,Amigo2012}. 
In this approach, given a measure-preserving map on a probability space, first 
an arbitrary finite partition of the space is taken. 
This gives rise to a finite-alphabet stationary stochastic process. 
An arbitrary ordering is introduced on the alphabet and the permutations of the words of finite lengths 
can be naturally defined (see Section 2 below). It is proved that 
the Shannon entropy of the occurrence of the permutations of a fixed length normalized by the length converges 
in the limit of the large length of the permutations. The quantity obtained is called the permutation entropy rate 
(also called metric permutation entropy) and is shown to be equal to the entropy rate of the process. 
By taking the limit of finer partitions of the measurable space, the permutation entropy rate of the 
measure-preserving map is defined if the limit exists. Amig\'{o} \cite{Amigo2012} proved that it exists and is equal to 
the Kolmogorov-Sinai entropy. 

In this paper, we restrict our attention to finite-alphabet stationary stochastic processes. 
Thus, we follow the second approach, namely, ordering on the alphabet is introduced arbitrarily. 
For quantities other than the entropy rate, 
three results for finite-alphabet stationary stochastic Markov processes 
have been shown by our previous work: the equality between the excess entropy and the permutation excess entropy \cite{Haruna2011a}, 
the equality between the mutual information expression of the excess entropy and its permutation analogue \cite{Haruna2011b} and 
the equality between the transfer entropy rate and the symbolic transfer entropy rate \cite{Haruna2011c}. 

The purpose of this paper is to set up a systematic framework to discuss permutation analogues of 
many information theoretic measures other than the entropy rate. 
In particular, we generalize our previous results for finite-alphabet stationary 
ergodic Markov processes to output processes of finite-state finite-alphabet hidden Markov models 
with ergodic internal processes. Upon this generalization, somewhat \textit{ad hoc} proofs in our previous work 
become systematic and greatly simplified. This makes us easily access quantities that have not been considered theoretically 
in the permutation approach. In this paper, we shall treat the following quantities: 
excess entropy \cite{Crutchfield2003}, transfer entropy \cite{Schreiber2000,Kaiser2002}, momentary information transfer \cite{Pompe2011} and directed information \cite{Marko1973,Massey1990}. 

This paper is organized as follows: 
In Section 2, we briefly review our previous result on the duality between words and permutations to make this paper as self-contained as possible. 
In Section 3, we prove a lemma about finite-state finite-alphabet hidden Markov models. 
In Section 4, we show equalities between various information theoretic complexity and coupling measures 
and their permutation analogues that hold for output processes of finite-state finite-alphabet hidden Markov models 
with ergodic internal processes. 
In Section 5, we discuss how our results are related to the previous work in the literature. 

\section{The duality between words and permutations}
In this section, we summarize the results from our previous work \cite{Haruna2011a} 
which will be used in this paper. 

Let $A_n$ be a finite set consisting of natural numbers from $1$ to $n$ called an \textit{alphabet}. 
In this paper $A_n$ is considered as a totally ordered set ordered by the usual `less-than-or-equal-to' relationship. 
When we emphasize the total order, we call $A_n$ \textit{ordered alphabet}. 

The set of all permutations of length $L \geq 1$ is denoted by $\mathcal{S}_L$. 
Namely, $\mathcal{S}_L$ is the set of all bijections $\pi$ on the set $\{1,2,\cdots,L\}$. 
For convenience, we sometimes denote a permutation $\pi$ of length $L$ 
by a string $\pi(1)\pi(2)\cdots\pi(L)$. 
The number of \textit{descents}, places with $\pi(i)>\pi(i+1)$, of $\pi \in \mathcal{S}_L$ 
is denoted by ${\rm Desc}(\pi)$. For example, if $\pi \in \mathcal{S}_5$ is given by 
$\pi(1)\pi(2)\pi(3)\pi(4)\pi(5)=35142$, then ${\rm Desc}(\pi)=2$. 

Let $A_n^L=\underbrace{A_n \times \cdots \times A_n}_{L}$ be the $L$-fold product of $A_n$. 
A \textit{word} of length $L \geq 1$ is an element of $A_n^L$. It is 
denoted by $x_{1:L}:=x_1 \cdots x_L:=(x_1,\cdots,x_L) \in A_n^L$. 
We say that the \textit{permutation type} of a word $x_{1:L}$ is $\pi \in \mathcal{S}_L$ 
if we have 
$x_{\pi(i)} \leq x_{\pi(i+1)}$ and $\pi(i) < \pi(i+1)$ when $x_{\pi(i)} = x_{\pi(i+1)}$ 
for $i =1,2,\cdots,L-1$. Namely, the permutation type of $x_{1:L}$ is 
the permutation of indices defined by re-ordering symbols $x_1,\cdots,x_L$ 
in the increasing order. 
For example, the permutation type of $x_{1:5}=31212 \in A_3^5$ is 
$\pi(1)\pi(2)\pi(3)\pi(4)\pi(5)=24351$ because $x_2 x_4 x_3 x_5 x_1=11223$. 

Let $\phi_{n,L}: A_n^L \to \mathcal{S}_L$ be a map sending each word $x_{1:L}$ 
to its permutation type $\pi=\phi_{n,L}(x_{1:L})$. 
We define another map $\mu_{n,L}: \phi_{n,L}\left( A_n^L \right) \subseteq \mathcal{S}_L \to A_n^L$ 
by the following procedure: 
\begin{itemize}
\item[(i)]
Given a permutation $\pi \in \phi_{n,L}\left( A_n^L \right) \subseteq \mathcal{S}_L$, 
we decompose the sequence $\pi(1) \cdots \pi(L)$ of length $L$ 
into \textit{maximal ascending subsequences}. 
A subsequence $i_j \cdots i_{j+k}$ of a sequence $i_1 \cdots i_L$ of length $L$ 
is called a \textit{maximal ascending subsequence} 
if it is ascending, namely, $i_j \leq i_{j+1} \leq \cdots \leq i_{j+k}$, 
and neither $i_{j-1} i_{j} \cdots i_{j+k}$ nor $i_{j} i_{j+1} \cdots i_{j+k+1}$ is ascending. 
\item[(ii)]
If $\pi(1) \cdots \pi(i_1), \ \pi(i_1+1) \cdots \pi(i_2), \cdots, \pi(i_{k-1}+1) \cdots \pi(L)$ 
is a decomposition of $\pi(1)\cdots\pi(L)$ into maximal ascending subsequences, then 
a word $x_{1:L} \in A_n^L$ is defined by 
\begin{eqnarray*}
x_{\pi(1)}=\cdots=x_{\pi(i_1)}=1, x_{\pi(i_1+1)}=\cdots=x_{\pi(i_2)}=2, \cdots, x_{\pi(i_{k-1})+1}=\cdots=x_{\pi(L)}=k. 
\end{eqnarray*}
We define $\mu_{n,L}(\pi)=x_{1:L}$. 
Note that ${\rm Desc}(\pi) \leq n-1$ because $\pi$ is the permutation type of some word $y_{1:L} \in A_n^L$. 
Thus, we have $k ={\rm Desc}(\pi)+1 \leq n$. 
Hence, $\mu_{n,L}$ is well-defined as a map from $\phi_{n,L}\left( A_n^L \right)$ to $A_n^L$. 
\end{itemize}

By construction, we have $\phi_{n,L} \circ \mu_{n,L}(\pi)=\pi$ for all $\pi \in \phi_{n,L}\left( A_n^L \right)$. 
To illustrate the construction of $\mu_{n,L}$, let us consider a word $y_{1:5}=21123 \in A_3^5$. 
The permutation type of $y_{1:5}$ is $\pi(1)\pi(2)\pi(3)\pi(4)\pi(5)=23145$. 
The decomposition of $23145$ into maximal ascending subsequences is 
$23,145$. We obtain $\mu_{n,L}(\pi)=x_1 x_2 x_3 x_4 x_5=21122$ by putting $x_2 x_3 x_1 x_4 x_5=11222$. 

\begin{Thm}
\begin{itemize}
\item[(i)]
For any $\pi \in {\mathcal S}_L$, 
\begin{equation*}
|\phi_{n,L}^{-1}(\pi)| = \binom{L+n-{\rm Desc}(\pi)-1}{L}, 
\end{equation*} 
where $\binom{a}{b}=0$ if $a<b$. 
\item[(ii)]
Let us put 
\begin{eqnarray*}
B_{n,L} &:=& \{x_{1:L} \in A_n^L | \phi_{n,L}^{-1}(\pi)=\{x_{1:L}\} \text{ for some } \pi \in \mathcal{S}_L \}, \\
C_{n,L} &:=& \{\pi \in \mathcal{S}_L | |\phi_{n,L}^{-1}(\pi)|=1 \}. 
\end{eqnarray*}
Then, $\phi_{n,L}$ restricted on $B_{n,L}$ is a map into $C_{n,L}$, 
$\mu_{n,L}$ restricted on $C_{n,L}$ is a map into $B_{n,L}$, 
and they form a pair of mutually inverse maps. Furthermore, we have 
\begin{eqnarray*}
B_{n,L} &=& \{x_{1:L} \in A_n^L | 1 \leq \forall i \leq n-1 \ 1 \leq \exists j < k \leq L \text{ s. t. } x_j=i+1,x_k=i \}, \\
C_{n,L} &=& \{\pi \in \mathcal{S}_L | {\rm Desc}(\pi)=n-1 \}. 
\end{eqnarray*}
\end{itemize}
\label{thm1}
\end{Thm}
{\it Proof.}
The theorem is a recasting of statements in Lemma 5 and Theorem 9 in \cite{Haruna2011a}. 
\hfill $\Box$ \\

Let ${\bf X}=\{X_1,X_2,\cdots\}$ be a finite-alphabet stationary stochastic process, 
where each stochastic variable $X_i$ takes its value in $A_n$. 
By the assumed stationarity, the probability of the occurrence of any word 
$x_{1:L} \in A_n^L$ is time-shift invariant: 
\begin{eqnarray*}
{\rm Pr}\{X_1=x_1,\cdots,X_L=x_L\}={\rm Pr}\{X_{k+1}=x_1,\cdots,X_{k+L}=x_L\}
\end{eqnarray*}
for any $k, L \geq 1$. Hence, it makes sense to define it without referring to 
the time to start. We denote the probability of the occurrence of a word 
$x_{1:L} \in A_n^L$ by $p(x_{1:L})=p(x_1 \cdots x_L)$. 
The probability of the occurrence of a permutation $\pi \in \mathcal{S}_L$ is 
given by $p(\pi)=\sum_{x_{1:L} \in \phi_{n,L}^{-1}(\pi)} p(x_{1:L})$. 

For a finite-alphabet stationary stochastic process ${\bf X}$ over the alphabet $A_n$, 
we define 
\begin{eqnarray*}
\alpha_{{\bf X}, L}:=\sum_{\begin{subarray}{c} \pi \in \mathcal{S}_L, \\ |\phi_{n,L}^{-1}(\pi)|>1 \end{subarray} } p(\pi) 
=\sum_{\pi \not\in C_{n,L} } p(\pi) 
\end{eqnarray*}
and
\begin{eqnarray*}
\beta_{x,{\bf X},L} &=& {\rm Pr}\{ x_{1:N} \in A_n^N | x_j \neq x \text{ for any } 1 \leq j \leq N \} \\
&=& \sum_{\begin{subarray}{c} x_j \neq x, \\ 1 \leq j \leq N \end{subarray} } p(x_1 \cdots x_N), 
\end{eqnarray*}
where $L \geq 1$, $x \in A_n$ and $N=\lfloor L/2 \rfloor$ and $\lfloor a \rfloor$ is the largest integer not greater than $a$. 

\begin{Lem}
Let ${\bf X}$ be a finite-alphabet stationary stochastic process and 
$\epsilon$ be a positive real number. 
If $\beta_{x,{\bf X},L} < \epsilon$ for any $x \in A_n$, 
then we have $\alpha_{{\bf X},L} < 2n \epsilon$. 
\label{lem2}
\end{Lem}
{\it Proof.}
The claim follows from Theorem \ref{thm1} (ii). See Lemma 12 in \cite{Haruna2011a} for the complete proof. 
\hfill $\Box$ \\

\section{A result on finite-state finite-alphabet hidden Markov models}

A \textit{finite-state finite-alphabet hidden Markov model} (in short, HMM) \cite{Upper1997} is a quadruple 
$(\Sigma,A,\{T^{(a)}\}_{a \in A},\mu)$, where $\Sigma$ and $A$ are finite sets 
called \textit{state set} and \textit{alphabet}, respectively, $\{T^{(a)}\}_{a \in A}$ is a family of 
$|\Sigma| \times |\Sigma|$ matrices indexed by elements of $A$ where $|\Sigma|$ is the size of state set $\Sigma$,
and $\mu$ is a probability distribution on the set $\Sigma$. 
The following conditions must be satisfied: 
\begin{itemize}
\item[(i)]
$T_{s s'}^{(a)} \geq 0$ for any $s,s' \in \Sigma$ and $a \in A$, 
\item[(ii)]
$\sum_{s',a} T_{s s'}^{(a)}=1$ for any $s \in \Sigma$, 
\item[(iii)]
and $\mu(s')=\sum_{s,a}\mu(s)T_{s s'}^{(a)}$ for any $s' \in \Sigma$. 
\end{itemize}
Any probability distribution satisfying the condition (iii) is called 
a \textit{stationary distribution}. 
The $|\Sigma| \times |\Sigma|$ matrix $T:=\sum_{a \in A} T^{(a)}$ 
is called \textit{state transition matrix}. The ternary $(\Sigma,T,\mu)$ 
defines the \textit{underlying Markov chain}. Note that the condition (iii) 
is equivalent to the condition (iii') $\mu(s')=\sum_{s}\mu(s)T_{s s'}$. 

Two finite-alphabet stationary processes are induced by a 
HMM $(\Sigma,A,\{T^{(a)}\}_{a \in A},\mu)$. One is solely determined by 
the underlying Markov chain. It is called 
\textit{internal process} and is denoted by ${\bf S}=\{S_1,S_2,\cdots\}$. 
The alphabet for ${\bf S}$ is $\Sigma$. 
The joint probability distributions which characterize ${\bf S}$ is given by 
\begin{equation*}
{\rm Pr}\{S_1=s_1,S_2=s_2,\cdots,S_L=s_L \}:=\mu(s_1)T_{s_1 s_2} \cdots T_{s_{L-1} s_L}
\end{equation*}
for any $s_1,\cdots,s_L \in \Sigma$ and $L \geq 1$. The other process 
${\bf X}=\{X_1,X_2,\cdots\}$ with the alphabet $A$ is defined by the following 
joint probability distributions 
\begin{equation*}
{\rm Pr}\{X_1=x_1,X_2=x_2,\cdots,X_L=x_L \}:=\sum_{s, s'}\mu(s)\left(T^{(x_1)} \cdots T^{(x_L)}\right)_{s s'}
\end{equation*}
for any $x_1,\cdots,x_L \in A$ and $L \geq 1$ and called \textit{output process}. 
The stationarity of the probability distribution $\mu$ ensures 
that of both the internal and output processes. 

Symbols $a \in A$ such that $T^{(a)}=O$ occur in the output process 
with probability $0$. Hence, we obtain the equivalent output process even if we remove these symbols. 
Thus, we can assume that $T^{(a)} \neq O$ for any $a \in A$ without loss of generality. 

The internal process ${\bf S}$ of a HMM $(\Sigma,A,\{T^{(a)}\}_{a \in A},\mu)$ 
is called \textit{ergodic} if the state transition matrix $T$ is \textit{irreducible} \cite{Walters1982}: 
for any $s,s' \in \Sigma$ there exists $k>0$ such that $(T^k)_{s s'}>0$. 
If the internal process ${\bf S}$ is ergodic, then the stationary distribution $\mu$ 
is uniquely determined by the state transition matrix $T$ via the condition (iii'). 
It is known that the ergodicity of the internal process ${\bf S}$ implies that of 
the output process ${\bf X}$, but not vice versa \cite{Lohr2010}. 

Note that there are two types of hidden Markov models depending on whether outputs are emitted from 
edges or states. The HMM defined here is edge emitting type. However, it is known that these 
two classes of HMM are equivalent \cite{Upper1997}. 
In particular, any finite-alphabet finite-order stationary Markov process can be described as a HMM defined here. 

\begin{Lem}
Let ${\bf X}$ be the output process of a HMM $(\Sigma,A_n,\{T^{(a)}\}_{a \in A_n},\mu)$, 
where $A_n=\{1,2,\cdots,n\}$ is an ordered alphabet. 
If the internal process ${\bf S}$ of the HMM is ergodic, then 
for any $x \in A_n$ 
there exists $0<\gamma_x<1$ and $C_x>0$ such that 
$\beta_{x,{\bf X},L}<C_x \gamma_x^L$ for any $L \geq 1$. 
\label{lem3}
\end{Lem}
{\it Proof.}
Given $L \geq 1$, let us put $N:=\lfloor L/2 \rfloor$. 
Fix any $x \in A_n$. We have 
\begin{eqnarray*}
\beta_{x,{\bf X},L} &=& \sum_{\begin{subarray}{c} x_j \neq x, \\ 1 \leq j \leq N \end{subarray} } p(x_1 \cdots x_N) \\
&=& \sum_{\begin{subarray}{c} x_j \neq x, \\ 1 \leq j \leq N \end{subarray} } \sum_{s, s'}\mu(s)\left(T^{(x_1)} \cdots T^{(x_N)}\right)_{s s'} \\
&=& \langle \mu \left( T - T^{(x)}\right)^N, {\bf 1} \rangle, 
\end{eqnarray*}
where ${\bf 1}=(1,1,\cdots,1)$ and $\langle \cdots,\cdots \rangle$ is the usual inner product 
of the $|\Sigma|$-dimensional Euclidean space $\mathbb{R}^{|\Sigma|}$. 
The largest eigenvalue $\lambda_x$ of the matrix $T_{(x)}:=T - T^{(x)}$ is less than 1. 
Indeed, this follows immediately from the Perron-Frobenius theorem for non-negative irreducible matrices: 
Let $M$ be a non-negative irreducible matrix. Then, 
\begin{itemize}
\item[(i)]
there exists a positive eigenvalue $\lambda_M$ called 
the \textit{Perron-Frobenius eigenvalue} such that any other eigenvalue of $M$ has absolute value not greater than $\lambda_M$, 
\item[(ii)]
there exists a positive eigenvector ${\bf v}$ corresponding to the eigenvalue $\lambda_M$, 
\item[(iii)]
and if $M \geq M' \geq O$ then $\lambda_M \geq \rho(M')$, where $\rho(M')$ is the spectral radius of non-negative matrix $M'$. 
The equality holds if and only if $M=M'$. 
\end{itemize}
Now, $T$ is a non-negative irreducible matrix with $\lambda_T=1$ by the assumption. 
Since $T \geq T_{(x)} \geq O$ and $T \neq T_{(x)}$, applying (iii) above implies that 
$\lambda_x=\rho(T_{(x)})<1$. 

We can show that for any $\epsilon>0$ there exists $C_\epsilon>0$ 
such that for any $k \geq 1$
\begin{equation*}
\| \mu T_{(x)}^k \| \leq C_{\epsilon} (\lambda_x + \epsilon)^k \| \mu \|, 
\end{equation*}
where $\| \cdots \|$ is the Euclidean norm and we used the fact that any non-negative matrix 
and its transpose have the same the Perron-Frobenius eigenvalue. 
For the proof of this inequality, see the beginning of section 1.2 in \cite{Katok1995}, for example. 
Since we have $\lambda_x<1$, we can choose $\epsilon>0$ so that $\lambda_x+\epsilon<1$. 
If we put $\gamma_x:=(\lambda_x+\epsilon)^\frac{1}{2}$ and 
$C_x:=C_{\epsilon} (\lambda_x+\epsilon)^{-1} \| \mu \| \| {\bf 1}\|$ then we obtain 
$\beta_{x,{\bf X},L}<C_x \gamma_x^L$ by the Cauchy-Schwartz inequality as desired. 

\hfill $\Box$ \\

\section{Permutation complexity and coupling measures}
In this section, we discuss the equalities between complexity and coupling measures and 
their permutation analogues for the output processes of HMMs whose internal processes are ergodic. 

\subsection{Fundamental lemma}
Let $({\bf X}^1,\cdots,{\bf X}^m)$ be a multivariate finite-alphabet stationary stochastic process, 
where each univariate process ${\bf X}^k=\{X_1^k,X_2^k,\cdots\}, \ k=1,2,\cdots,m$ is defined 
over an ordered alphabet $A_{n_k}$. For simplicity, we use the notations 
\begin{eqnarray*}
p(x_{a_1:b_1}^1,\cdots,x_{a_m:b_m}^m)
&:=&{\rm Pr}\{ X_{a_1:b_1}^1 = x_{a_1:b_1}^1,\cdots, X_{a_m:b_m}^m = x_{a_m:b_m}^m \},\\
p(\pi_1,\cdots,\pi_m)
&:=&{\rm Pr}\{ \phi_{n_k,b_k-a_k+1} \circ X_{a_k:b_k}^k = \pi_k, \ k=1,\cdots,m \} 
\end{eqnarray*}
and 
\begin{eqnarray*}
p(\pi_k)
:={\rm Pr}\{ \phi_{n_k,b_k-a_k+1} \circ X_{a_k:b_k}^k = \pi_k\}, 
\end{eqnarray*}
where $1 \leq a_k \leq b_k$, $x_{a_k:b_k}^k \in A_{n_k}^{b_k-a_k+1}$ 
and $\pi_k \in \mathcal{S}_{b_k-a_k+1}$ for $k=1,\cdots,m$. 

\begin{Lem}
\begin{eqnarray*}
0 &\leq& H(X_{a_1:b_1}^1,\cdots,X_{a_m:b_m}^m)-H^*(X_{a_1:b_1}^1,\cdots,X_{a_m:b_m}^m) \\
&\leq& \left( \sum_{k=1}^m \alpha_{{\bf X}^k,b_k-a_k+1} \right) \left( \sum_{k=1}^m n_k \log(b_k-a_k+1+n_k) \right), 
\end{eqnarray*}
where 
\begin{eqnarray*}
H(X_{a_1:b_1}^1,\cdots,X_{a_m:b_m}^m)= - \sum_{x_{a_1:b_1}^1,\cdots, x_{a_m:b_m}^m} p(x_{a_1:b_1}^1,\cdots, x_{a_m:b_m}^m) \log p(x_{a_1:b_1}^1,\cdots, x_{a_m:b_m}^m)
\end{eqnarray*}
and 
\begin{eqnarray*}
H^*(X_{a_1:b_1}^1,\cdots,X_{a_m:b_m}^m)= - \sum_{\pi_1,\cdots, \pi_m} p(\pi_1,\cdots,\pi_m) \log p(\pi_1,\cdots,\pi_m) 
\end{eqnarray*}
are the Shannon entropy of the joint occurrence of 
words $x_{a_1:b_1}^1,\cdots, x_{a_m:b_m}^m$ and permutations $\pi_1,\cdots,\pi_m$, respectively, 
and the base of the logarithm is taken as $2$. 
\label{lem4}
\end{Lem}
{\it Proof.}
We have 
\begin{eqnarray*}
&& H(X_{a_1:b_1}^1,\cdots,X_{a_m:b_m}^m)-H^*(X_{a_1:b_1}^1,\cdots,X_{a_m:b_m}^m) \\
&& = \sum_{\begin{subarray}{c} \pi_1,\cdots,\pi_m, \\ p(\pi_1,\cdots,\pi_m)>0 \end{subarray} } p(\pi_1,\cdots,\pi_m) \\
&& \times \left( -\sum_{ \begin{subarray}{c} x_{a_k:b_k}^k \in \phi_{n_k,b_k-a_k+1}^{-1}(\pi_k), \\ 1 \leq k \leq m \end{subarray} } \frac{p(x_{a_1:b_1}^1,\cdots, x_{a_m:b_m}^m)}{p(\pi_1,\cdots,\pi_m)} \log \frac{p(x_{a_1:b_1}^1,\cdots, x_{a_m:b_m}^m)}{p(\pi_1,\cdots,\pi_m)}
\right).
\end{eqnarray*}
By Theorem \ref{thm1} (i), it holds that 
\begin{eqnarray*}
0 &\leq& -\sum_{ \begin{subarray}{c} x_{a_k:b_k}^k \in \phi_{n_k,b_k-a_k+1}^{-1}(\pi_k), \\ 1 \leq k \leq m \end{subarray} } \frac{p(x_{a_1:b_1}^1,\cdots, x_{a_m:b_m}^m)}{p(\pi_1,\cdots,\pi_m)} \log \frac{p(x_{a_1:b_1}^1,\cdots, x_{a_m:b_m}^m)}{p(\pi_1,\cdots,\pi_m)} \\
&\leq& \log \left( \prod_{k=1}^m \binom{b_k - a_k + n_k - {\rm Desc}(\pi_k)}{b_k - a_k + 1} \right) \\
&\leq& \log \left( \prod_{k=1}^m (b_k - a_k + 1 + n_k)^{n_k} \right) \\
&=& \sum_{k=1}^m n_k \log (b_k - a_k + 1 + n_k)
\end{eqnarray*}
for $(\pi_1,\cdots,\pi_m) \in \mathcal{S}_{b_1-a_1+1} \times \cdots \times \mathcal{S}_{b_m-a_m+1}$ 
such that $p(\pi_1,\cdots,\pi_m)>0$. 

If $|( \phi_{n_1,b_1-a_1+1} \times \cdots \phi_{n_m,b_m-a_m+1} )^{-1}(\pi_1,\cdots,\pi_m)|=1$ then 
\begin{eqnarray*}
-\sum_{ \begin{subarray}{c} x_{a_k:b_k}^k \in \phi_{n_k,b_k-a_k+1}^{-1}(\pi_k), \\ 1 \leq k \leq m \end{subarray} } \frac{p(x_{a_1:b_1}^1,\cdots, x_{a_m:b_m}^m)}{p(\pi_1,\cdots,\pi_m)} \log \frac{p(x_{a_1:b_1}^1,\cdots, x_{a_m:b_m}^m)}{p(\pi_1,\cdots,\pi_m)}
=0.
\end{eqnarray*}

On the other hand, we have 
\begin{eqnarray*}
\sum_{\begin{subarray}{c} \pi_1,\cdots,\pi_m, \\ \exists k \text{ s.t. } |\phi_{n_k,b_k-a_k+1}^{-1}(\pi_k)|>1 \end{subarray} } p(\pi_1,\cdots,\pi_m) 
&\leq& \sum_{k=1}^m 
\sum_{\begin{subarray}{c} \pi_k, \\ |\phi_{n_k,b_k-a_k+1}^{-1}(\pi_k)|>1 \end{subarray} } p(\pi_k) \\
&=& \sum_{k=1}^m \alpha_{{\bf X}^k, b_k-a_k+1}. 
\end{eqnarray*}
This completes the proof of the inequality. 

\hfill $\Box$ \\

\subsection{Excess Entropy}
Let ${\bf X}$ be a finite-alphabet stationary stochastic process. 
Its \textit{excess entropy} is defined by \cite{Crutchfield2003}
\begin{eqnarray*}
{\bf E}({\bf X})
&=& \lim_{L \to \infty} \left( H(X_{1:L}) - h({\bf X})L \right) \\
&=& \sum_{L=1}^{\infty} \left( H(X_L|X_{1:L-1}) - h({\bf X}) \right), 
\end{eqnarray*}
if the limit on the right-hand side exists, where $h({\bf X})=\lim_{L \to \infty}H(X_{1:L})/L$ 
is the \textit{entropy rate} of ${\bf X}$ which exists 
for any finite-alphabet stationary stochastic process \cite{Cover1991}. 

The excess entropy has been used as a measure of complexity \cite{Arnold1996,Bialek2001,Feldman2008,Grassberger1986,Li1991,Shaw1984}. 
Actually, it quantifies global correlations present in a given stationary process in the following sense. 
If ${\bf E}({\bf X})$ exists then it can be written as 
the mutual information between the past and future: 
\begin{eqnarray*}
{\bf E}({\bf X})=\lim_{L \to \infty} I(X_{1:L};X_{L+1:2L}). 
\end{eqnarray*}
It is known that if ${\bf X}$ is the output process of a HMM then ${\bf E}({\bf X})$ exists \cite{Lohr2010}. 

When the alphabet of ${\bf X}$ is an ordered alphabet $A_n$, we define 
the \textit{permutation excess entropy} of ${\bf X}$ by \cite{Haruna2011a}
\begin{eqnarray*}
{\bf E}^*({\bf X})
&=& \lim_{L \to \infty} \left( H^*(X_{1:L}) - h^*({\bf X})L \right) \\
&=& \sum_{L=1}^{\infty} \left( H^*(X_L|X_{1:L-1}) - h^*({\bf X}) \right), 
\end{eqnarray*}
if the limit on the right-hand side exists, where $h^*({\bf X})=\lim_{L \to \infty}H^*(X_{1:L})/L$ 
is the \textit{permutation entropy rate} of ${\bf X}$ which exists 
for any finite-alphabet stationary stochastic process and is equal to the entropy rate $h({\bf X})$ 
\cite{Amigo2005,Amigo2010,Amigo2012} and $H^*(X_L|X_{1:L-1}):=H^*(X_{1:L})-H^*(X_{1:L-1})$. 

The following proposition is a generalization of our previous results in \cite{Haruna2011a,Haruna2011b}. 
\begin{Pro}
Let ${\bf X}$ be the output process of a HMM $(\Sigma,A_n,\{T^{(a)}\}_{a \in A_n},\mu)$ 
with an ergodic internal process. Then, we have 
\begin{equation*}
{\bf E}({\bf X})={\bf E}^*({\bf X})=\lim_{L \to \infty} I^*(X_{1:L};X_{L+1:2L}), 
\end{equation*}
where $I^*(X_{1:L};X_{L+1:2L}):=H^*(X_{1:L})+H^*(X_{L+1:2L})-H^*(X_{1:L},X_{L+1:2L})=2H^*(X_{1:L})-H^*(X_{1:L},X_{L+1:2L})$. 
\label{pro5}
\end{Pro}
{\it Proof.}
Let $L \geq 1$. We have 
\begin{eqnarray*}
&&| \left( H(X_{1:L}) - h({\bf X})L \right) - \left( H^*(X_{1:L}) - h^*({\bf X})L \right)| \\
&=& |H(X_{1:L}) - H^*(X_{1:L})| \\
&\leq& \alpha_{{\bf X},L} n \log(L+n) \\
&\leq& 2 C n^2 \log (L+n) \gamma^L, 
\end{eqnarray*}
where $C:=\max_{x \in A_n} \{ C_x\}$, $\gamma:=\max_{x \in A_n} \{ \gamma_x \}<1$ and 
we have used $h({\bf X})=h^*({\bf X})$ for the first equality, Lemma \ref{lem4} for the second inequality 
and Lemma \ref{lem2} and Lemma \ref{lem3} for the last inequality. 
By taking the limit $L \to \infty$ we obtain ${\bf E}({\bf X})={\bf E}^*({\bf X})$. 

To prove $\lim_{L \to \infty} I(X_{1:L};X_{L+1:2L})=\lim_{L \to \infty} I^*(X_{1:L};X_{L+1:2L})$, 
it is sufficient to show that $| H(X_{1:L},X_{L+1:2L}) - H^*(X_{1:L},X_{L+1:2L})| \to 0$ as 
$L \to \infty$. This is because we have 
\begin{eqnarray*}
&&|I(X_{1:L};X_{L+1:2L}) - I^*(X_{1:L};X_{L+1:2L})| \\
&\leq& 2| H(X_{1:L}) - H(X_{1:L})| + | H(X_{1:L},X_{L+1:2L}) - H^*(X_{1:L},X_{L+1:2L})|. 
\end{eqnarray*}
However, this can be shown similarly with the above discussion 
by applying Lemma \ref{lem4} to the bivariate process $({\bf X}^1,{\bf X}^2):=({\bf X},{\bf X})$ 
and then using Lemma \ref{lem2} and Lemma \ref{lem3}. 

\hfill $\Box$ \\

\subsection{Transfer Entropy and Momentary Information Transfer}

In this subsection we consider two information rates that are measures of 
coupling direction and strength between two jointly distributed processes 
and discuss the equalities between them and their permutation analogues. 
One is the rate of the transfer entropy \cite{Schreiber2000} 
and the other is the rate of the momentary information transfer \cite{Pompe2011}. 
Both are particular instances of the conditional mutual information \cite{Frenzel2007}. 

Let $({\bf X},{\bf Y})$ be a bivariate finite-alphabet stationary stochastic process. 
We assume that the alphabets of ${\bf X}$ and ${\bf Y}$ are ordered alphabets $A_n$ and $A_m$, 
respectively. 
For $\tau=1,2,\cdots$, we define the \textit{$\tau$-step transfer entropy rate} from ${\bf Y}$ to ${\bf X}$ by 
\begin{eqnarray*}
t_\tau({\bf Y} \to {\bf X}) 
= \lim_{L \to \infty} && \big[ H(X_{1:L+\tau}) - H(X_{1:L}) \\
&& - H(X_{1:L+\tau},Y_{1:L}) + H(X_{1:L},Y_{1:L}) \big]. 
\end{eqnarray*}
When $\tau=1$, $t_1({\bf Y} \to {\bf X})$ is called just \textit{transfer entropy rate} \cite{Amblard2011} 
from ${\bf Y}$ to ${\bf X}$ and simply denoted by $t({\bf Y} \to {\bf X})$. 

If we introduce the \textit{$\tau$-step entropy rate} of ${\bf X}$ by 
\begin{equation*}
h_\tau({\bf X})=\lim_{L \to \infty} H(X_{L+1:L+\tau}|X_{1:L}) 
\end{equation*}
and the \textit{$\tau$-step conditional entropy rate} of ${\bf X}$ given ${\bf Y}$ by 
\begin{equation*}
h_\tau({\bf X} | {\bf Y})=\lim_{L \to \infty} H(X_{L+1:L+\tau}|X_{1:L},Y_{1:L}) 
\end{equation*}
then we can write 
\begin{equation*}
t_\tau({\bf Y} \to {\bf X}) = h_\tau ({\bf X}) - h_\tau ({\bf X} | {\bf Y}) 
\end{equation*}
because both $h_\tau ({\bf X})$ and $h_\tau ({\bf X} | {\bf Y})$ exist. 
We call $h_1({\bf X} | {\bf Y})$ \textit{conditional entropy rate} and denote it by 
$h({\bf X} | {\bf Y})$\footnote{
Note that the conditional entropy rate here is slightly different from that found in the literature. 
For example, in \cite{Ash1965}, conditional entropy rate (called \textit{conditional uncertainty}) 
is defined by $\lim_{L \to \infty} H(X_{L+1}|X_{1:L},Y_{1:L+1})$. 
The difference from the conditional entropy rate defined here is in whether the conditioning on $Y_{L+1}$ is involved or not. 
}. 

$h_\tau({\bf X})$ is \textit{additive}, namely, 
we always have 
\begin{equation*}
h_\tau({\bf X})=\tau h_1({\bf X})=\tau h({\bf X}). 
\end{equation*}
However, for the $\tau$-step conditional entropy rate, the additivity cannot hold in general. 
It is at most \textit{super-additive}: 
we only have the inequality 
\begin{equation*}
h_\tau ({\bf X} | {\bf Y}) \geq \tau h({\bf X} | {\bf Y}) 
\end{equation*}
in general. Indeed, we have 
\begin{eqnarray*}
h_\tau({\bf X} | {\bf Y}) &=& \lim_{L \to \infty} H(X_{L+1:L+\tau}|X_{1:L},Y_{1:L}) \\
&=& \lim_{L \to \infty} \sum_{\tau'=1}^\tau H(X_{L+\tau'}|X_{1:L+\tau'-1},Y_{1:L}) \\
&\geq& \lim_{L \to \infty} \sum_{\tau'=1}^\tau H(X_{L+\tau'}|X_{1:L+\tau'-1},Y_{1:L+\tau'-1}) \\
&=& \tau h({\bf X} | {\bf Y}). 
\end{eqnarray*}
This leads to the \textit{sub-additivity} of the $\tau$-step transfer entropy rate: 
\begin{equation*}
t_\tau ({\bf Y} \to {\bf X}) \leq \tau t({\bf Y} \to {\bf X}). 
\end{equation*}
An example with the strict inequality can be easily given. Let ${\bf Y}$ 
be an i.i.d. process and ${\bf X}$ be defined by $X_1=Y_1$ and $X_{i+1}=Y_i$. 
We have $h({\bf X})=h({\bf Y})=H(Y_1)$ and $h_\tau({\bf X} | {\bf Y})=(\tau -1) H(Y_1)$. 
Hence, $t_\tau ({\bf Y} \to {\bf X})=H(Y_1)$ for any $\tau=1,2,\cdots$. 

There are two permutation analogues of the transfer entropy. 
One is called the \textit{symbolic transfer entropy (STE)} \cite{Staniek2008} and the other is called 
the \textit{transfer entropy on rank vector (TERV)} \cite{Kugiumtzis2012}. Here, we introduce their rates as follows: 
the \textit{rate of STE} from ${\bf Y}$ to ${\bf X}$ is defined by 
\begin{eqnarray*}
t_\tau^{**}({\bf Y} \to {\bf X}) 
= \lim_{L \to \infty} && \big[ H^*(X_{1:L},X_{1+\tau:L+\tau}) - H^*(X_{1:L}) \\
&& - H^*(X_{1:L},X_{1+\tau:L+\tau},Y_{1:L}) + H^*(X_{1:L},Y_{1:L}) \big] 
\end{eqnarray*}
if the limit on the right-hand side exists. 
The \textit{rate of TERV} from ${\bf Y}$ to ${\bf X}$ is defined by 
\begin{eqnarray*}
t_\tau^{*}({\bf Y} \to {\bf X}) 
= \lim_{L \to \infty} && \big[ H^*(X_{1:L+\tau}) - H^*(X_{1:L}) \\
&& - H^*(X_{1:L+\tau},Y_{1:L}) + H^*(X_{1:L},Y_{1:L}) \big]. 
\end{eqnarray*}
if the limit on the right-hand side exists. 
If ${\bf E}^*({\bf X})$ exists then, by the definition of the permutation excess entropy, we have 
\begin{eqnarray*}
h^*({\bf X})=\lim_{L \to \infty} \left( H^*(X_{1:L+1}) - H^*(X_{1:L}) \right). 
\end{eqnarray*}
In this case, $t_1^{*}({\bf Y} \to {\bf X})$ coincides with a quantity called \textit{symbolic transfer entropy rate} 
introduced in \cite{Haruna2011c}. 

\begin{Pro}
Let $({\bf X},{\bf Y})$ be the output process of a HMM 
$(\Sigma,A_n \times A_m,\{T^{(a,b)}\}_{(a,b) \in A_n \times A_m},\mu)$ 
with an ergodic internal process. Then, we have 
\begin{equation*}
t_\tau({\bf Y} \to {\bf X})=t_\tau^{*}({\bf Y} \to {\bf X})=t_\tau^{**}({\bf Y} \to {\bf X}). 
\end{equation*}
\label{pro6}
\end{Pro}
{\it Proof.}
Since both ${\bf X}$ and ${\bf Y}$ are the output processes of appropriate HMMs with 
ergodic internal processes, the equalities follow from the similar discussion with that 
in the proof of Proposition \ref{pro5}. Indeed, for example, ${\bf X}$ is the output process of 
the HMM $(\Sigma,A_n,\{T^{(a)}\}_{a \in A_n},\mu)$ where $T^{(a)}:=\sum_{b \in A_m}T^{(a,b)}$. 

\hfill $\Box$ \\

A different instance of conditional mutual information called \textit{momentary information transfer} 
is considered in \cite{Pompe2011}. It was proposed to improve the ability to detect coupling delays which 
is lacked in the transfer entropy. Here, we consider its rate: 
the \textit{momentary information transfer rate} is defined by 
\begin{eqnarray*}
m_\tau({\bf Y} \to {\bf X}) 
= \lim_{L \to \infty} && \big[ H(X_{1:L+\tau},Y_{1:L-1}) - H(X_{1:L+\tau-1},Y_{1:L-1}) \\
&& - H(X_{1:L+\tau},Y_{1:L}) + H(X_{1:L+\tau-1},Y_{1:L}) \big]. 
\end{eqnarray*}
Its permutation analogue called \textit{momentary sorting information transfer rate} is defined by 
\begin{eqnarray*}
m_\tau^*({\bf Y} \to {\bf X}) 
= \lim_{L \to \infty} && \big[ H^*(X_{1:L+\tau},Y_{1:L-1}) - H^*(X_{1:L+\tau-1},Y_{1:L-1}) \\
&& - H^*(X_{1:L+\tau},Y_{1:L}) + H^*(X_{1:L+\tau-1},Y_{1:L}) \big]. 
\end{eqnarray*}
By the similar discussion with that in the proof of Proposition \ref{pro6}, we obtain the following 
equality: 

\begin{Pro}
Let $({\bf X},{\bf Y})$ be the output process of a HMM 
$(\Sigma,A_n \times A_m,\{T^{(a,b)}\}_{(a,b) \in A_n \times A_m},\mu)$ 
with an ergodic internal process. Then, we have 
\begin{equation*}
m_\tau({\bf Y} \to {\bf X})=m_\tau^{*}({\bf Y} \to {\bf X}). 
\end{equation*}
\label{pro7}
\end{Pro}

\subsection{Directed Information}

\textit{Directed information} is a measure of coupling direction and strength based on the idea 
of the \textit{causal conditioning} \cite{Massey1990,Kramer1998}. 
Since it is not a particular instance of conditional mutual information, here we treat it separately. 
In the following presentation, we make use of terminologies from \cite{Amblard2011,Amblard2012}. 

Let $({\bf X},{\bf Y})$ be a bivariate finite-alphabet stationary stochastic process. 
The alphabets of ${\bf X}$ and ${\bf Y}$ are ordered alphabets $A_n$ and $A_m$, respectively. 
The \textit{directed information rate} from ${\bf Y}$ to ${\bf X}$ is defined by 
\begin{eqnarray*}
I_\infty ({\bf Y} \to {\bf X}) 
= \lim_{L \to \infty} \frac{1}{L} I(Y_{1:L} \to X_{1:L}) 
\end{eqnarray*}
where 
\begin{eqnarray*}
I(Y_{1:L} \to X_{1:L}) 
&=& \sum_{i=1}^L I(X_i,Y_{1:i}|X_{1:i-1}) \\
&=& H(X_{1:L}) - \sum_{i=1}^L H(X_i | X_{1:i-1},Y_{1:i}). 
\end{eqnarray*}
Note that if $Y_{1:i}$ in the above expression on the right-hand side is replaced by 
$Y_{1:L}$ then we obtain the mutual information between $X_{1:L}$ and $Y_{1:L}$: 
\begin{eqnarray*}
I(X_{1:L};Y_{1:L}) = H(X_{1:L}) - \sum_{i=1}^L H(X_i | X_{1:i-1},Y_{1:L}). 
\end{eqnarray*}
Thus, conditioning on $Y_{1:i}$ for $i=1,\cdots,L$, not on $Y_{1:L}$, 
distinguishes the directed information from the mutual information. 
Following \cite{Kramer1998}, we write 
\begin{eqnarray*}
H(X_{1:L} || Y_{1:L}) := \sum_{i=1}^L H(X_i | X_{1:i-1},Y_{1:i}) 
\end{eqnarray*}
and call the quantity \textit{causal conditional entropy}. By using this notation, we have 
\begin{equation*}
I(Y_{1:L} \to X_{1:L}) = H(X_{1:L}) - H(X_{1:L} || Y_{1:L}). 
\end{equation*}

The permutation analogue of the directed information rate which we call 
\textit{symbolic directed information rate} is defined by 
\begin{eqnarray*}
I_\infty^* ({\bf Y} \to {\bf X}) 
= \lim_{L \to \infty} \frac{1}{L} I^*(Y_{1:L} \to X_{1:L}) 
\end{eqnarray*}
if the limit on the right-hand side exists, where 
\begin{eqnarray*}
I^*(Y_{1:L} \to X_{1:L}) := H^*(X_{1:L}) - \sum_{i=1}^L \left( H^*(X_{1:i},Y_{1:i}) - H^*(X_{1:i-1},Y_{1:i}) \right). 
\end{eqnarray*}
If we write 
\begin{equation*}
I^*(X_i;Y_{1:i} | X_{1:i-1}):=H^*(X_{1:i})-H^*(X_{1:i-1})-H^*(X_{1:i},Y_{1:i})+H^*(X_{1:i-1},Y_{1:i})
\end{equation*}
and 
\begin{eqnarray*}
H^*(X_{1:L} || Y_{1:L}) := \sum_{i=1}^L \left( H^*(X_{1:i},Y_{1:i}) - H^*(X_{1:i-1},Y_{1:i}) \right)
\end{eqnarray*}
then we have the expressions 
\begin{equation*}
I^*(Y_{1:L} \to X_{1:L}) = \sum_{i=1}^L I^*(X_i;Y_{1:i} | X_{1:i-1}) = H^*(X_{1:L}) - H^*(X_{1:L} || Y_{1:L}). 
\end{equation*}

\begin{Pro}
Let $({\bf X},{\bf Y})$ be the output process of a HMM 
$(\Sigma,A_n \times A_m,\{T^{(a,b)}\}_{(a,b) \in A_n \times A_m},\mu)$ 
with an ergodic internal process. Then, we have 
\begin{equation*}
I_\infty ({\bf Y} \to {\bf X})=I_\infty^{*}({\bf Y} \to {\bf X}). 
\end{equation*}
\label{pro8}
\end{Pro}
{\it Proof.}
We have 
\begin{eqnarray*}
&& | I(Y_{1:L} \to X_{1:L}) - I^*(Y_{1:L} \to X_{1:L}) | \\
&&\leq |H(X_{1:L}) - H^*(X_{1:L})| + \sum_{i=1}^L |H(X_{1:i},Y_{1:i}) - H^*(X_{1:i},Y_{1:i})| \\
&& \hspace{4cm} + \sum_{i=1}^L |H(X_{1:i-1},Y_{1:i}) - H^*(X_{1:i-1},Y_{1:i})|. 
\end{eqnarray*}
We know that the first term on the right-hand side in the above inequality goes to $0$ as 
$L \to \infty$. Let us evaluate the second sum. By Lemma \ref{lem4}, it holds that 
\begin{eqnarray*}
\sum_{i=1}^L |H(X_{1:i},Y_{1:i}) - H^*(X_{1:i},Y_{1:i})| 
\leq \sum_{i=1}^L \left( \alpha_{{\bf X},i} + \alpha_{{\bf Y},i} \right) \left( n \log(i+n) + m \log(i+m) \right)
\end{eqnarray*}
By Lemma \ref{lem2} and Lemma \ref{lem3}, we have 
\begin{eqnarray*}
\sum_{i=1}^L \alpha_{{\bf X},i} n \log(i+n) \leq 2 C n^2 \sum_{i=1}^L \gamma^i \log (i+n), 
\end{eqnarray*}
where $C:=\max_{x \in A_n} \{ C_x\}$ and $\gamma:=\max_{x \in A_n} \{ \gamma_x \}<1$. 
It is elementary to show that $\lim_{L \to \infty} \sum_{i=1}^L \gamma^i \log (i+n)$ is finite. 
The limits of the other terms are also shown to be finite similarly. 
Thus, we can conclude that the limit of the second sum is bounded. 
Similarly, the limit of the third sum is also bounded. The equality in the claim follows immediately. 

\hfill $\Box$ \\

For output processes of HMMs with ergodic internal processes, properties 
on the directed information rate can be transferred to those on the symbolic directed information rate. 
Since proofs of them can be given by the same manner as those of the above propositions, 
here we list some of them without proofs. For the proofs of the properties on the directed information rate, 
we refer to \cite{Amblard2012,Kramer1998}. 

Let $({\bf X},{\bf Y})$ be the output process of a HMM 
$(\Sigma,A_n \times A_m,\{T^{(a,b)}\}_{(a,b) \in A_n \times A_m},\mu)$ 
with an ergodic internal process. Then, we have 
\begin{itemize}
\item[(i)]
\begin{equation*}
I_\infty^* ({\bf Y} \to {\bf X})=\lim_{L \to \infty} I^{*}(X_L;Y_{1:L} | X_{1:L-1}). 
\end{equation*}
This is the permutation analogue of the equality 
\begin{equation*}
I_\infty({\bf Y} \to {\bf X})=\lim_{L \to \infty} I(X_L;Y_{1:L} | X_{1:L-1}). 
\end{equation*}

\item[(ii)]
\begin{eqnarray*}
I_\infty(D{\bf Y} \to {\bf X})=I_\infty^*(D{\bf Y} \to {\bf X})=\lim_{L \to \infty} I^*(X_L;Y_{1:L-1}|X_{1:L-1}). 
\end{eqnarray*}
Here, 
\begin{eqnarray*}
I_\infty(D{\bf Y} \to {\bf X}):=\lim_{L \to \infty} \frac{1}{L} I(DY_{1:L} \to X_{1:L}) 
\end{eqnarray*}
and 
\begin{eqnarray*}
I(DY_{1:L} \to X_{1:L}) := \sum_{i=1}^L I(X_i ; Y_{1:i-1} | X_{1:i-1}). 
\end{eqnarray*}
The symbol $D$ denotes the one-step delay. $I_\infty^*(D{\bf Y} \to {\bf X})$ is the corresponding permutation analogue. 
The second equality is the permutation analogue of the equality 
$I_\infty(D{\bf Y} \to {\bf X})=\lim_{L \to \infty} I(X_L;Y_{1:L-1}|X_{1:L-1})$. 
Since $I_\infty(D{\bf Y} \to {\bf X})$ coincides with the transfer entropy rate, 
the first equality is just the equality between the transfer entropy rate and the 
symbolic transfer entropy rate (or the rate of 1-step TERV) proved in Proposition \ref{pro6} 
given the second equality. 

\item[(iii)]
\begin{eqnarray*}
I_\infty({\bf Y} \to {\bf X} || D{\bf Y})=I_\infty^*({\bf Y} \to {\bf X} || D{\bf Y})=\lim_{L \to \infty} I^*(X_L;Y_L|X_{1:L-1},Y_{1:L-1}), 
\end{eqnarray*}
where $I_\infty({\bf Y} \to {\bf X} || D{\bf Y})$ is called the \textit{instantaneous information exchange rate} and 
is defined by 
\begin{eqnarray*}
I_\infty({\bf Y} \to {\bf X} || D{\bf Y}) := \lim_{L \to \infty} \frac{1}{L} I(Y_{1:L} \to X_{1:L} || DY_{1:L}) 
\end{eqnarray*}
and 
\begin{eqnarray*}
I(Y_{1:L} \to X_{1:L} || DY_{1:L}) &=& H(X_{1:L} || DY_{1:L}) - H(X_{1:L} || Y_{1:L}, DY_{1:L}) \\
&=& \sum_{i=1}^L I(X_i;Y_{1:i} | X_{1:i-1}, Y_{1:i-1}) \\
&=& \sum_{i=1}^L I(X_i;Y_i | X_{1:i-1}, Y_{1:i-1}). 
\end{eqnarray*}
From the last expression of $I(Y_{1:L} \to X_{1:L} || DY_{1:L})$, we can obtain 
\begin{eqnarray*}
I_\infty({\bf Y} \to {\bf X} || D{\bf Y})=\lim_{L \to \infty} I(X_L;Y_L | X_{1:L-1}, Y_{1:L-1}). 
\end{eqnarray*}
$I_\infty^*({\bf Y} \to {\bf X} || D{\bf Y})$ is the corresponding permutation analogue and 
called \textit{symbolic instantaneous information exchange rate}. 

\item[(iv)]
\begin{eqnarray*}
I_\infty^*({\bf Y} \to {\bf X})=I_\infty^*(D{\bf Y} \to {\bf X})+I_\infty^*({\bf Y} \to {\bf X} || D{\bf Y}). 
\end{eqnarray*}
Namely, the symbolic directed information rate decomposes into the sum of 
the symbolic transfer entropy rate and the symbolic instantaneous information exchange rate. 
This follows immediately from (ii), (iii) and the equality saying that 
the directed information rate decomposes into the sum of the transfer entropy rate and the instantaneous 
information exchange rate: 
\begin{eqnarray*}
I_\infty({\bf Y} \to {\bf X})=I_\infty(D{\bf Y} \to {\bf X})+I_\infty({\bf Y} \to {\bf X} || D{\bf Y}). 
\end{eqnarray*}

\item[(v)]
\begin{eqnarray*}
I_\infty^*({\bf Y} \to {\bf X}) + I_\infty^*(D{\bf X} \to {\bf Y}) = I_\infty^*({\bf X};{\bf Y}). 
\end{eqnarray*}
This is the permutation analogue of the equality saying that 
the mutual information rate between ${\bf X}$ and ${\bf Y}$ is the sum of 
the directed information rate from ${\bf Y}$ to ${\bf X}$ and the transfer entropy rate 
from ${\bf X}$ to ${\bf Y}$: 
\begin{eqnarray*}
I_\infty({\bf Y} \to {\bf X}) + I_\infty(D{\bf X} \to {\bf Y}) = I_\infty({\bf X};{\bf Y}), 
\end{eqnarray*}
where 
\begin{eqnarray*}
I_\infty({\bf X};{\bf Y}) := \lim_{L \to \infty} \frac{1}{L} I(X_{1:L} ; Y_{1:L})
\end{eqnarray*}
is the \textit{mutual information rate} and $I_\infty^*({\bf X};{\bf Y})$ is its permutation analogue 
called \textit{symbolic mutual information rate}. 
It is known that they are equal for any bivariate finite-alphabet stationary stochastic process \cite{Haruna2011c}. 
Thus, the symbolic mutual information rate between ${\bf X}$ and ${\bf Y}$ is the sum of 
the symbolic directed information rate from ${\bf Y}$ to ${\bf X}$ and the symbolic transfer entropy rate 
from ${\bf X}$ to ${\bf Y}$. 
\end{itemize}

We can also introduce the permutation analogue of the \textit{causal conditional directed information rate} and 
prove the corresponding properties. To be precise, let us consider a multivariate finite-alphabet 
stationary stochastic process $({\bf X},{\bf Y},{\bf Z}^1,\cdots,{\bf Z}^k)$ with the alphabet 
$A_n \times A_m \times A_{l_1} \times \cdots \times A_{l_k}$. 
The \textit{causal conditional directed information rate} from ${\bf Y}$ to ${\bf X}$ given 
$({\bf Z}^1,\cdots,{\bf Z}^k)$ is defined by 
\begin{eqnarray*}
I_\infty({\bf Y} \to {\bf X} || {\bf Z}^1,\cdots,{\bf Z}^k) := \lim_{L \to \infty} \frac{1}{L} I(Y_{1:L} \to X_{1:L} || Z_{1:L}^1,\cdots,Z_{1:L}^k) 
\end{eqnarray*}
where 
\begin{eqnarray*}
I(Y_{1:L} \to X_{1:L} || Z_{1:L}^1,\cdots,Z_{1:L}^k) &=& H(X_{1:L} || Z_{1:L}^1,\cdots,Z_{1:L}^k) - H(X_{1:L} || Y_{1:L},Z_{1:L}^1,\cdots,Z_{1:L}^k) \\
&=& \sum_{i=1}^L I(X_i ; Y_{1:i} | X_{1:i-1},Z_{1:L}^1,\cdots,Z_{1:L}^k). 
\end{eqnarray*}

Corresponding to Proposition \ref{pro8}, we have the following equality if 
$({\bf X},{\bf Y},{\bf Z}^1,\cdots,{\bf Z}^k)$ is the output process of a HMM with an ergodic internal 
process: 
\begin{eqnarray*}
I_\infty({\bf Y} \to {\bf X} || {\bf Z}^1,\cdots,{\bf Z}^k) = I_\infty^*({\bf Y} \to {\bf X} || {\bf Z}^1,\cdots,{\bf Z}^k), 
\end{eqnarray*}
where $I_\infty^*({\bf Y} \to {\bf X} || {\bf Z}^1,\cdots,{\bf Z}^k)$ is the 
\textit{symbolic causal conditional directed information rate} which is defined by the same manner as 
the symbolic directed information rate. The following properties also hold: 
assume that $({\bf X},{\bf Y},{\bf Z}^1,\cdots,{\bf Z}^k)$ is the output process of a HMM 
with an ergodic internal process. Then, we have 
\begin{itemize}
\item[(i')]
\begin{equation*}
I_\infty^* ({\bf Y} \to {\bf X} || {\bf Z}^1,\cdots,{\bf Z}^k)=\lim_{L \to \infty} I^{*}(X_L;Y_{1:L} | X_{1:L-1},Z_{1:L}^1,\cdots,Z_{1:L}^k). 
\end{equation*}
This is the permutation analogue of the equality 
\begin{equation*}
I_\infty({\bf Y} \to {\bf X} || {\bf Z}^1,\cdots,{\bf Z}^k)=\lim_{L \to \infty} I(X_L;Y_{1:L} | X_{1:L-1},Z_{1:L}^1,\cdots,Z_{1:L}^k). 
\end{equation*}

\item[(ii')]
\begin{eqnarray*}
I_\infty(D{\bf Y} \to {\bf X}  || {\bf Z}^1,\cdots,{\bf Z}^k)
&=& I_\infty^*(D{\bf Y} \to {\bf X}  || {\bf Z}^1,\cdots,{\bf Z}^k) \\
&=& \lim_{L \to \infty} I^*(X_L;Y_{1:L-1}|X_{1:L-1},Z_{1:L}^1,\cdots,Z_{1:L}^k). 
\end{eqnarray*}
The second equality is the permutation analogue of the equality 
\begin{eqnarray*}
I_\infty(D{\bf Y} \to {\bf X}  || {\bf Z}^1,\cdots,{\bf Z}^k)=\lim_{L \to \infty} I(X_L;Y_{1:L-1}|X_{1:L-1},Z_{1:L}^1,\cdots,Z_{1:L}^k). 
\end{eqnarray*}
The quantities $I_\infty(D{\bf Y} \to {\bf X}  || {\bf Z}^1,\cdots,{\bf Z}^k)$ and 
$I_\infty^*(D{\bf Y} \to {\bf X}  || {\bf Z}^1,\cdots,{\bf Z}^k)$ are called 
\textit{causal conditional transfer entropy rate} and 
\textit{symbolic causal conditional transfer entropy rate}, respectively. 

\item[(iii')]
\begin{eqnarray*}
I_\infty({\bf Y} \to {\bf X} || D{\bf Y},{\bf Z}^1,\cdots,{\bf Z}^k)
&=& I_\infty^*({\bf Y} \to {\bf X} || D{\bf Y},{\bf Z}^1,\cdots,{\bf Z}^k) \\
&=& \lim_{L \to \infty} I^*(X_L;Y_L|X_{1:L-1},Y_{1:L-1},Z_{1:L}^1,\cdots,Z_{1:L}^k), 
\end{eqnarray*}
where $I_\infty({\bf Y} \to {\bf X} || D{\bf Y},{\bf Z}^1,\cdots,{\bf Z}^k)$ is called \textit{causal conditional instantaneous information exchange rate}. 
The second equality is the permutation analogue of the equality 
\begin{eqnarray*}
I_\infty({\bf Y} \to {\bf X} || D{\bf Y},{\bf Z}^1,\cdots,{\bf Z}^k)=\lim_{L \to \infty} I(X_L;Y_L | X_{1:L-1}, Y_{1:L-1},Z_{1:L}^1,\cdots,Z_{1:L}^k). 
\end{eqnarray*}
$I_\infty^*({\bf Y} \to {\bf X} || D{\bf Y},{\bf Z}^1,\cdots,{\bf Z}^k)$ is the permutation analogue and 
is called \textit{symbolic causal conditional instantaneous information exchange rate}. 

\item[(iv')]
\begin{eqnarray*}
I_\infty^*({\bf Y} \to {\bf X} || {\bf Z}^1,\cdots,{\bf Z}^k)=I_\infty^*(D{\bf Y} \to {\bf X} || {\bf Z}^1,\cdots,{\bf Z}^k)+I_\infty^*({\bf Y} \to {\bf X} || D{\bf Y},{\bf Z}^1,\cdots,{\bf Z}^k). 
\end{eqnarray*}
This is the permutation analogue of the following equality 
\begin{eqnarray*}
I_\infty({\bf Y} \to {\bf X} || {\bf Z}^1,\cdots,{\bf Z}^k)=I_\infty(D{\bf Y} \to {\bf X} || {\bf Z}^1,\cdots,{\bf Z}^k)+I_\infty({\bf Y} \to {\bf X} || D{\bf Y},{\bf Z}^1,\cdots,{\bf Z}^k). 
\end{eqnarray*}

\item[(v')]
\begin{eqnarray*}
I_\infty^*({\bf Y} \to {\bf X} || {\bf Z}^1,\cdots,{\bf Z}^k) + I_\infty^*(D{\bf X} \to {\bf Y} || {\bf Z}^1,\cdots,{\bf Z}^k) = I_\infty^*({\bf X};{\bf Y} || {\bf Z}^1,\cdots,{\bf Z}^k). 
\end{eqnarray*}
This is the permutation analogue of the equality 
\begin{eqnarray*}
I_\infty({\bf Y} \to {\bf X} || {\bf Z}^1,\cdots,{\bf Z}^k) + I_\infty(D{\bf X} \to {\bf Y} || {\bf Z}^1,\cdots,{\bf Z}^k) = I_\infty({\bf X};{\bf Y} || {\bf Z}^1,\cdots,{\bf Z}^k), 
\end{eqnarray*}
where 
\begin{eqnarray*}
I_\infty({\bf X};{\bf Y} || {\bf Z}^1,\cdots,{\bf Z}^k) &:=& \lim_{L \to \infty} \frac{1}{L} \big( H(X_{1:L} || Z_{1:L}^1,\cdots,Z_{1:L}^k) \\
&&+ H(Y_{1:L} || Z_{1:L}^1,\cdots,Z_{1:L}^k) - H(X_{1:L},Y_{1:L} || Z_{1:L}^1,\cdots,Z_{1:L}^k) \big)
\end{eqnarray*}
is the \textit{causal conditional mutual information rate} and $I_\infty^*({\bf X};{\bf Y} || {\bf Z}^1,\cdots,{\bf Z}^k)$ 
is its permutation analogue called \textit{symbolic causal conditional mutual information rate}. 
It can be shown that 
\begin{eqnarray*}
I_\infty({\bf X};{\bf Y} || {\bf Z}^1,\cdots,{\bf Z}^k)=I_\infty^*({\bf X};{\bf Y} || {\bf Z}^1,\cdots,{\bf Z}^k)
\end{eqnarray*}
if $({\bf X},{\bf Y},{\bf Z}^1,\cdots,{\bf Z}^k)$ is the output process of a HMM with an ergodic internal process. 
\end{itemize}

\section{Discussion}
In this section, we discuss how our theoretical results in this paper are related to the previous work in the literature. 

Being confronted with real time series data, we cannot take the limit of large length of words. 
Hence, we have to estimate information rates with finite length of words. In such situation, 
one permutation method could have some advantages to the other permutation methods. 
As a matter of fact, TERV was originally proposed as an improved analogue of STE \cite{Kugiumtzis2012}. 
However, it has been unclear whether they coincide in the limit of large length of permutations. 
In this paper, we provide a partial answer to this question: 
the two permutation analogues of the transfer entropy rate, the rate of STE and the rate of TERV, 
are equivalent to the transfer entropy rate for bivariate processes generated by HMMs with ergodic internal processes. 

Granger causality graph \cite{Dahlhaus2003} is a model of causal dependence structure in 
multivariate stationary stochastic processes. Given a multivariate stationary stochastic process, 
nodes in a Granger causality graph are components of the process. There are two types of edges: 
one is directed and the other is undirected. The absence of a directed edge from one node to 
another node indicates the lack of the Granger cause from the former to the latter relative to 
the other remaining processes. Similarly, the absence of a undirected edge between two nodes 
indicates the lack of the instantaneous cause between them relative to the other remaining processes. 
Amblard and Michel \cite{Amblard2011,Amblard2012} proposed that the Granger causality graph can be constructed 
based on the directed information theory: let ${\mathcal X}=({\bf X}^1,{\bf X}^2,\cdots,{\bf X}^m)$ be 
a multivariate finite-alphabet stationary stochastic process with the alphabet 
$A_{n_1} \times A_{n_2} \times \cdots \times A_{n_m}$ and $(V,E_d,E_u)$ be the Granger causality graph 
of the process ${\mathcal X}$ where $V=\{1,2,\cdots,m\}$ is the set of nodes, $E_d$ is the set of 
directed edges and $E_u$ is the set of undirected edges. Their proposal is that 
\begin{itemize}
\item[(i)]
for any $i,j \in V$, $(i,j) \not \in E_d$ if and only if 
$I_\infty(D{\bf X}^i \to {\bf X}^j || {\mathcal X} \setminus \{{\bf X}^i,{\bf X}^j\}) = 0$, 
\item[(ii)]
for any $i,j \in V$, $(i,j) \not \in E_u$ if and only if 
$I_\infty({\bf X}^i \to {\bf X}^j || D{\bf X}^i,{\mathcal X} \setminus \{{\bf X}^i,{\bf X}^j\}) = 0$. 
\end{itemize}
Thus, in the Granger causality graph construction proposed in \cite{Amblard2011}, 
the causal conditional transfer entropy rate captures the Granger cause from one process 
to another process relative to the other remaining processes. On the other hand, the causal conditional 
instantaneous information exchange rate captures the instantaneous cause between two processes relative to the 
other remaining processes. 

Now, let us consider the case when ${\mathcal X}$ is an output process of a HMM with an ergodic internal process. 
Then, from the results of Section 4.4, we have 
\begin{itemize}
\item[(i')]
for any $i,j \in V$, $(i,j) \not \in E_d$ if and only if 
$I_\infty^*(D{\bf X}^i \to {\bf X}^j || {\mathcal X} \setminus \{{\bf X}^i,{\bf X}^j\}) = 0$, 
\item[(ii')]
for any $i,j \in V$, $(i,j) \not \in E_u$ if and only if 
$I_\infty^*({\bf X}^i \to {\bf X}^j || D{\bf X}^i,{\mathcal X} \setminus \{{\bf X}^i,{\bf X}^j\}) = 0$. 
\end{itemize}
Thus, the Granger causality graphs in the sense of \cite{Amblard2011,Amblard2012} for multivariate processes generated by 
HMMs with ergodic internal processes can be captured by the language of the permutation entropy: 
the symbolic causal conditional transfer entropy rate and the symbolic instantaneous information exchange rate. 
This statement opens up a possibility of the permutation approach to the problem of assessing 
the causal dependence structure of multivariate stationary stochastic processes. 
However, of course, the details of the practical implementation should be an issue of further study. 

\subsection*{Acknowledgments}
The authors would like to thank D. Kugiumtzis for his useful comments and discussion 
on the relationship between STE and TERV. TH was supported by the JST PRESTO program. 

\bibliographystyle{plain}
\bibliography{phmm_v2}

\end{document}